\documentclass[conference]{IEEEtran}
\IEEEoverridecommandlockouts
\usepackage{cite}
\usepackage{amsmath,amssymb,amsfonts}
\usepackage{algorithmic}
\usepackage{graphicx}
\usepackage{textcomp}
\usepackage{xcolor}
\usepackage{subfig}
\usepackage{hyperref}
\def\BibTeX{{\rm B\kern-.05em{\sc i\kern-.025em b}\kern-.08em
    T\kern-.1667em\lower.7ex\hbox{E}\kern-.125emX}}

\begin{document}

\title{Transformer Network for Semantically-Aware and Speech-Driven Upper-Face Generation
\thanks{This work was performed within the Labex SMART (ANR-11-LABX-65) supported by French state funds managed by the ANR within the Investissements d'Avenir programme under reference ANR-11-IDEX-0004-02.}}

\author{\IEEEauthorblockN{Mireille Fares}
\IEEEauthorblockA{\textit{ISIR, STMS} \\
\textit{Sorbonne Université}\\}
\and
\IEEEauthorblockN{Catherine Pelachaud}
\IEEEauthorblockA{\textit{CNRS} \\
\textit{ISIR, Sorbonne Université}}
\and
\IEEEauthorblockN{Nicolas Obin}
\IEEEauthorblockA{\textit{STMS} \\
\textit{IRCAM, Sorbonne Université, CNRS}}
}
\maketitle
\vspace*{-1.2cm}
\begin{abstract} We propose a semantically-aware speech driven model to generate expressive and natural upper-facial and head motion for Embodied Conversational Agents (ECA). In this work, we aim to produce natural and continuous head motion and upper-facial gestures synchronized with speech. We propose a model that generates these gestures based on multimodal input features: the first modality is text, and the second one is speech prosody. Our model makes use of Transformers and Convolutions to map the multimodal features that correspond to an utterance to continuous eyebrows and head gestures. We conduct subjective and objective evaluations to validate our approach and compare it with state of the art.\end{abstract}

\vspace{-0.4cm}
\section{Introduction}
\subsection{Context}
The human face is a key channel of communication in human-human interaction. During speech, humans spontaneously and continuously display various facial and head gestures that convey a large panel of information to the interlocutors. These gestures are known as ‘‘visual prosody'' \cite{visualprosody_Graf2002}: facial or head movement are produced in conjunction with verbal communication. 
During speech, the Fundamental Frequency (F0) variations and upper-facial movements are highly correlated \cite{yehia2002linking}; they are the results of linguistic and conversational choices. Eyebrow motion can also occur during pauses. Another kinematic–acoustic relation that happens during production of speech is between F0 and head motion \cite{yehia2002linking}. Natural and rhythmic head movements are one of the key factors in producing natural animations \cite{ding2013modeling, chen2020talking}.

\subsection{Related works and our contributions}
A large number of head motion generation systems has been proposed in previous works \cite{hofer2007automatic, haag2016bidirectional, lu2020prediction, vougioukas2019realistic, mariooryad2012generating, sadoughi2019speech, wang2021audio2head}. A variety of generative statistical models aimed to predict the multimodal behavior of a virtual agent. Hidden Markov Models (HMM) \cite{ hofer2007automatic}, Recurrent Neural Networks (RNN) \cite{wang2021audio2head, haag2016bidirectional}, and Dynamic Bayesian Networks (DBN) \cite{mariooryad2012generating, sadoughi2019speech} have been used to generate head motion from speech; Generative Adversarial Networks (GAN) have been proposed to produce facial gestures from speech \cite{karras2017audio,  vougioukas2019realistic}. \cite{kucherenko2020gesticulator} generates continuous 3D hand gestures based on acoustics and semantics. However, most of the aforementioned approaches exploit as input one modality only, namely speech, and neglect to render their approach using semantic information. Also, most of them focus only on facial expressions while the correlation between facial expressions and head movements are crucial to produce a natural behavior. For instance, \cite{ hofer2007automatic, haag2016bidirectional, lu2020prediction, sadoughi2019speech} do not generate eyebrow motion along with head motion, which are both correlated to F0 \cite{yehia2002linking}, and therefore are correlated to each other. On the other hand, transformer networks and attention mechanisms have been recently proved to be very efficient for sequence-to-sequence modelling, with particular advances for modelling  multimodal processes. For instance, Transformers were previously used for translating speech to text (ASR)\cite{hrinchuk2020correction, mohamed2019transformers}, and multimodal learning of images based on text \cite{yao2020multimodal}. To overcome those limitations, we propose a novel approach for upper-facial and head gestures generation based on a multimodal transformer network. 
Our contributions can be listed as follows: 1) a transformer network operating on multi-modal input text and speech information in order to generate upper-facial and head movements, and 2) a cross-attention module that can efficiently exploit semantic and speech information.
\footnote{Video samples of our model's gestures predictions and other related material can be found in: \url{https://github.com/mireillefares/VAAnimation/blob/main/README.md}}

The paper is organised as follows. The next section describes the proposed architecture of our model and its multi-modal features. Then our objective and subjective evaluation experiments are presented including the LSTM-based Baseline Model.  We finally discuss our results.
\section{Proposed architecture}
\subsection{Multimodal Input/Output Features}
The upper-facial movements and head rotations are represented by mean of Action Units (AUs) as defined in the Facial Action Coding Systems (FACS)\cite{ekman} and 3D head angles. The work presented in this paper only considers the AUs that represent eyebrows movements which are: inner raise eyebrow \emph{\textbf{AU1}}, outer raise eyebrow \emph{\textbf{AU2}}, frown \emph{\textbf{AU4}}, upper
lid raiser \emph{\textbf{AU5}}, cheek raiser \emph{\textbf{AU6}}, and lid tightener \emph{\textbf{AU7}}. Head rotations have three degrees of freedom, represented by the Euler angles \emph{\textbf{roll}}, \emph{\textbf{pitch}} and \emph{\textbf{yaw}}. They are represented by \emph{\textbf{RX}}, \emph{\textbf{RY}} and \emph{\textbf{RZ}} which are the rotation of the head with respect to the \emph{\textbf{X}}, \emph{\textbf{Y}} and \emph{\textbf{Z}} axes. 
In this paper, F0 values were extracted from the speech signal at a 5ms audio frame rate, linearly interpolated between unvoiced segments, and clipped to the range of 50 to 550Hz which represents the F0 range of human speech. Since F0, AU intensities (AU) and head rotations (R) are continuous, they were quantized to produce a finite set of discrete values to reduce the model size and energy consumption \cite{quantization}. Finally, BERT word embeddings were extracted from the text transcription. 
\begin{figure}[htb]
    \centering
    \subfloat[][\centering Model Architecture]{{\includegraphics[width=7.5cm]{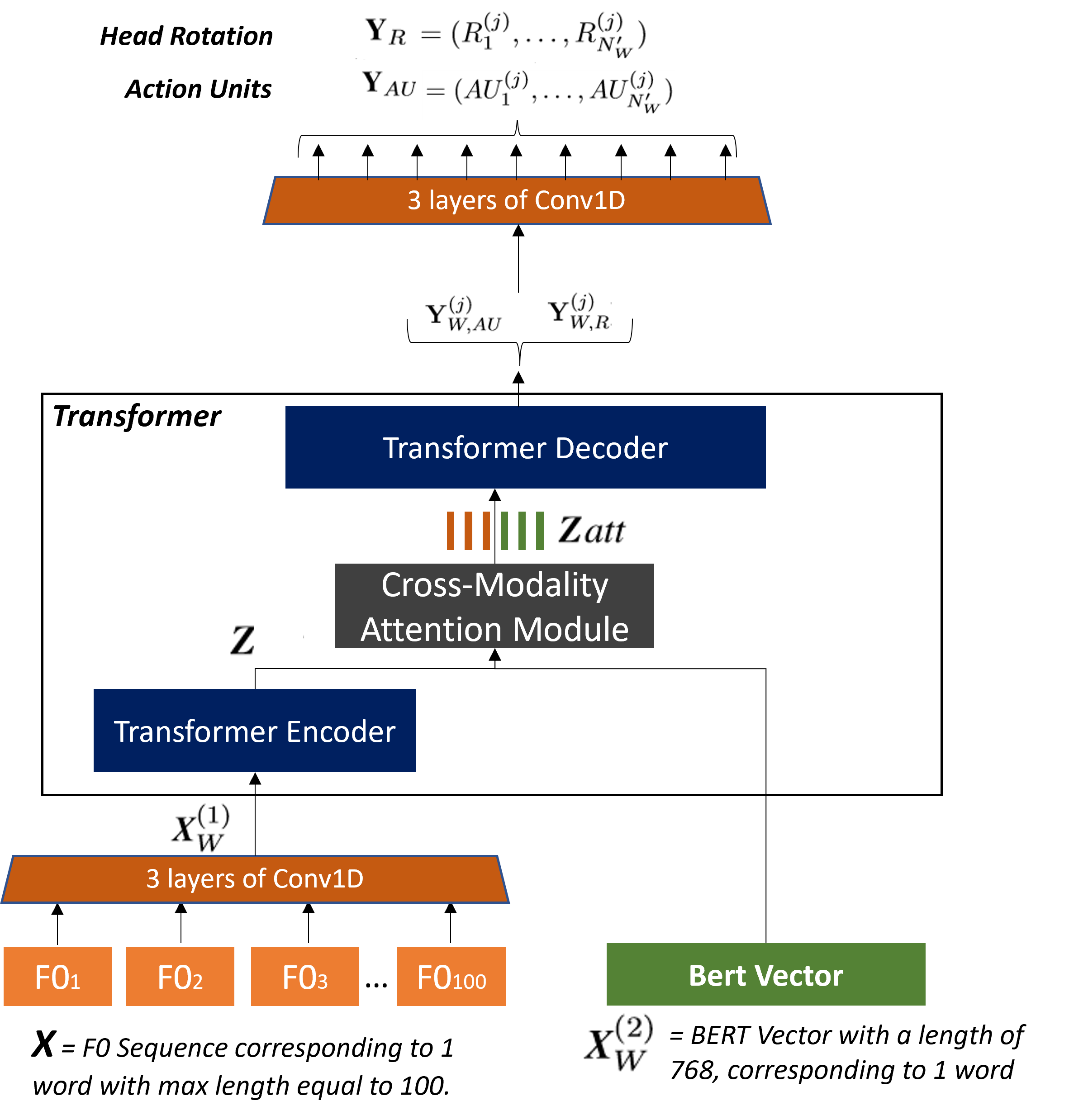}}}%
    \qquad
    \centering
    \subfloat[][\centering Transformer Encoder]{{\includegraphics[width=3.5cm]{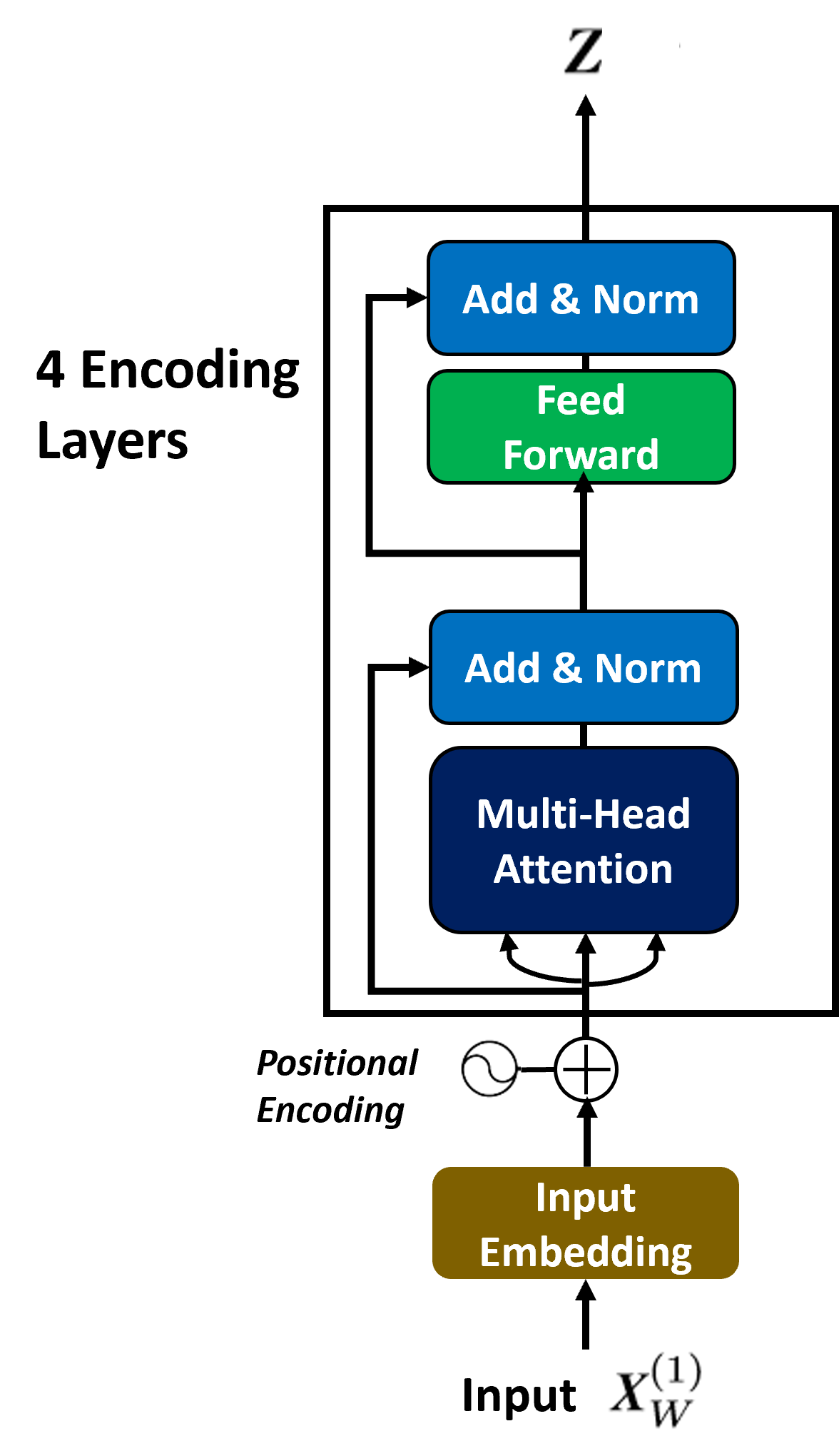} }}%
    \qquad
    \subfloat[][\centering Transformer Decoder]{{\includegraphics[width=4cm]{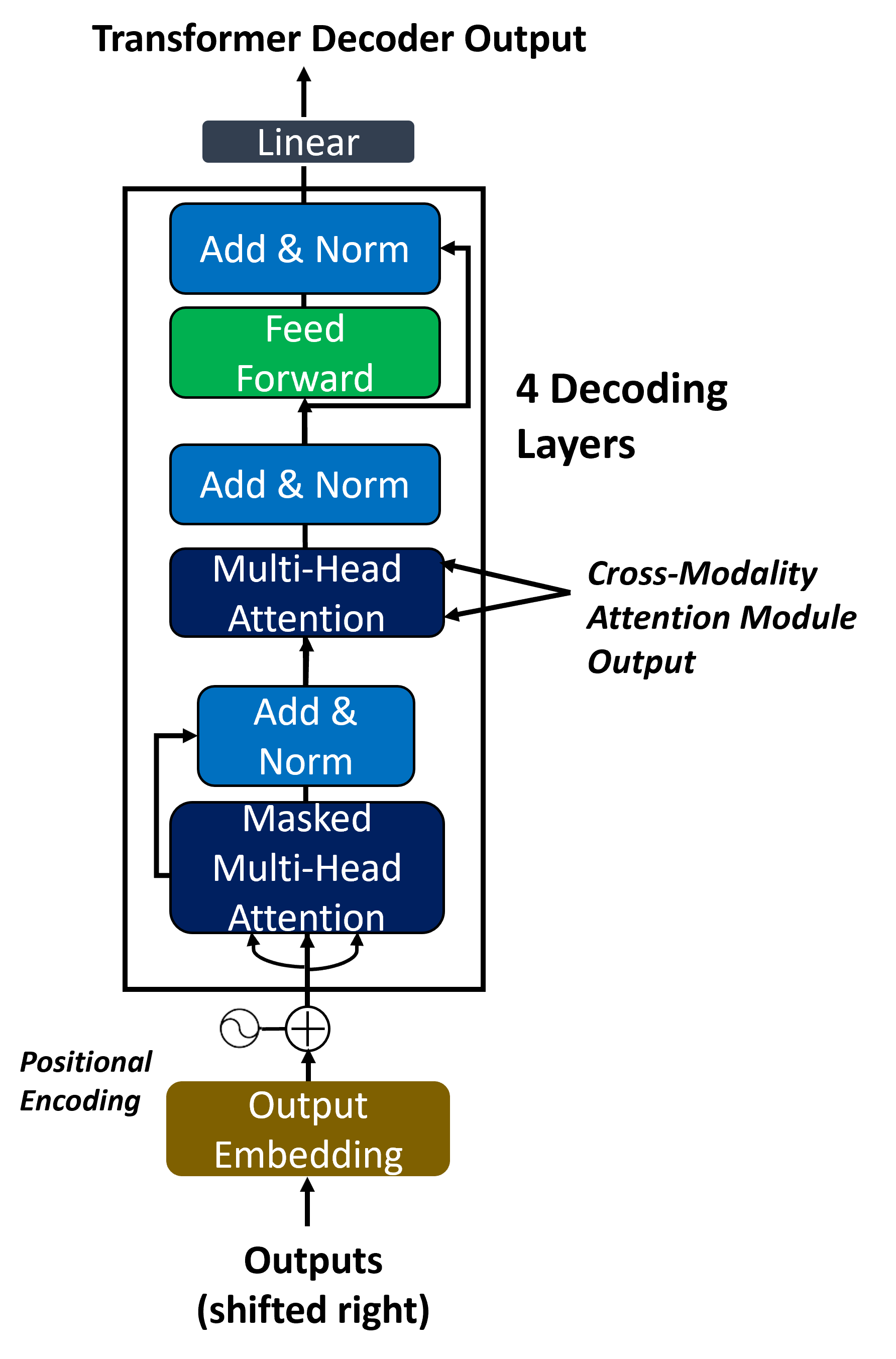} }}%
    \caption{Face Gesture Generation Model Architecture}%
    \label{fig:ModelArchitecture}%
    \vspace{-0.4cm}
\end{figure}

\subsection{Face Gesture Generation Model Architecture}
The proposed architecture aims at mapping multimodal speech and text feature sequence into continuous facial and head gestures. This problem is treated as a multimodal sequence-to-sequence problem, for which a transformer network operating at the word level is presented as illustrated in Fig.\ref{fig:ModelArchitecture} (a). The inputs and outputs of the transformer network consist of one feature vector for each word \emph{\textbf{W}} of the input text sequence. The input text corresponds to an Inter-Pausal Unit (IPU) which is a sequence of words separated by silent pauses longer than 0.2 seconds. F0 input sequence as well as AU intensities \emph{\textbf{AU}} and head rotations \emph{\textbf{R}} output sequences have a variable length. We set the maximum F0 input sequence length to 100, and the maximum \emph{\textbf{AU/R}} output length to 124. Shorter sequences were padded to the maximum length, and longer ones were truncated.
In order to handle continuous flow of input and output information with different timing, the transformer is wrapped with a F0 encoder module at its input and to a \emph{\textbf{AU/R}} decoder at its output. The objective of the F0 encoder is to encode the continuous F0-values at the word level and the objective of the \emph{\textbf{AU/R}} decoder is to reconstruct the continuous values for \emph{\textbf{AU}} and \emph{\textbf{R}} from a word-level encoding. Each spoken word \emph{\textbf{W}} is represented by a word-level F0 embedding vector  \emph{\textbf{X}$^{(1)}_W$} which at its turn corresponds to the sequence of F0-values ${\textbf{X}^{(1)}}=(F0_{1}, \dots, F0_{N_{W}})$, where $N_{W}$ is the number of F0-values corresponding to the spoken word \emph{\textbf{W}}; and \emph{\textbf{X}$^{(2)}_W$} is the BERT word embedding vector corresponding to the spoken word \emph{\textbf{W}} - including silences. Silences less than 0.2 secs may belong to IPUs. They do not have a contextual BERT embedding. Hence, we replaced them by a comma - \emph{\textbf{","}}. Each spoken word \emph{\textbf{W}} is also represented by a word-level \emph{\textbf{AU}} vector ${\mathbf{Y}^{(j)}_{W, {AU}}}$ (resp. \emph{\textbf{R}} vector ${\mathbf{Y}^{(j)}_{{W}, {R}}}$) which in turn corresponds to the sequence of values d${\textbf{Y}^{(j)}_{{AU}}= (AU_{1}^{(j)}, \dots, AU_{N'_W}^{(j)})}$ (resp. ${\textbf{Y}_{R}^{(j)}= (R_{1}^{(j)}, \dots, R_{N'_W}^{(j)})}$)   where $j$ denotes the $j^{th}$ \emph{\textbf{AU/R}} and $N'_W$ the number \emph{\textbf{AU/R}}-values corresponding to the word \emph{\textbf{W}}. %

\textbf{F0 Encoder}: As depicted in Fig.\ref{fig:ModelArchitecture} (a), for each \emph{\textbf{W}}, three one-dimensional convolutional layers are applied to project the input F0 sequence \emph{\textbf{X}$^{(1)}_W$} into a word-level representation of F0 contours covering local context of F0 variations. These convolutional layers include \emph{64} filters, with a \emph{kernel size} equal to \emph{3}. The generated output vector of the latter layers \emph{\textbf{X}$^{(1)}_W$} is then fed as an input to the \emph{Transformer Encoder}.  

\textbf{Transformer Encoder}: The transformer encoder architecture is depicted in Fig.\ref{fig:ModelArchitecture} (b); it is similar to the one proposed in \cite{vaswani2017attention}. In our work, it is composed of a stack of \emph{N = 4} identical layers. Each layer has two sub-layers: the first one is a multi-head self attention mechanism with \emph{4 attention heads}, and the second one is a position-wise fully connected feed-forward network. As the original transformer encoder, we employ a residual connection around each of the 2 sub-layers, followed by layer normalization.%

\textbf{Cross-Modality Attention Module (CMAM)}: The output of the transformer encoder \emph{\textbf{Z}}, as well as \emph{\textbf{X}$^{(2)}_W$} are fed as inputs to the Cross-Modality Attention Module (see Fig.\ref{fig:ModelArchitecture} (a)). This Module has the same structure as the Transformer Decoder in \cite{vaswani2017attention}. It generates a representation that can take into account both modalities, text and speech. The representation learning is done in a master/slave manner, where one modality - the master - is used to highlight the extracted features in  the other modality - the slave. This module takes \emph{\textbf{X}$^{(2)}_W$} - Text Modality - as master, and \emph{\textbf{Z}} - Speech Modality - as  slave. Thus, it performs cross-attention such that the attention mask is derived from text modality, and is harnessed to leverage the latent features from the speech modality.

\textbf{Transformer Decoder}: The decoder is composed of \emph{N = 4} identical decoding layers, with \emph{4 attention heads}. Similar to the one proposed in \cite{vaswani2017attention}, it is composed of residual connections applied around each of the sub-layers, followed by layer normalization. As depicted in Fig.\ref{fig:ModelArchitecture} (c), the self-attention sub-layer in the decoder stack is modified to prevent positions from attending to subsequent positions. The output predictions which are offset by one position, and this masking ensure that the predictions for position index \emph{\textbf{j}} depend only on the known outputs at positions less than \emph{\textbf{j}}. We use 6 Transformer Decoders, one for each \emph{\textbf{AU/R}}. For simplicity, Fig.\ref{fig:ModelArchitecture} (a) only illustrates one decoder. 
 
\textbf{AU/R decoder}: As depicted in Fig.\ref{fig:ModelArchitecture} (a), the Transformer Decoder outputs are concatenated together, then fed to 3 one-dimensional convolutional layers that include 64 filters, with a kernel size equal to 3, to learn the correlation between the 6 output features, and therefore the correlation between facial and head movements. Finally, a Dense layer with a Softmax activation function is applied on each of the outputs, to convert the outputs to predicted next-token probabilities. The final output sequences are ${\mathbf{Y}_{{AU}}}$ and ${\mathbf{Y}_{{R}}}$.

\textbf{Transformer Sub-Layers and Hyperparameters}: the transformer encoder and decoders have attention sub-layers, and contain fully connected feed-forward networks which are applied to each position separately and identically. Similarly to other sequence to sequence models, we use learned embeddings to convert the input tokens and output tokens to vectors of dimension d$_{model}$ = 64. All sub-layers and embedding layers therefore use this dimension. The inner feed-forward layers are of dimension d$_{ff}$ = 400. Positional encodings are applied to the inputs of the transformer encoder and decoders. They have the same dimension as the embeddings, so that they can be added together. We use sine and cosine functions, similar to \cite{vaswani2017attention}.  

\section{Experiments}
\subsection{Material and Experimental Setups}
We trained it on a subset of the TED dataset collected in \cite{fares2020towards}, containing preprocessed AUs/R, F0s, and BERT embeddings of filtered shots where speakers' face and head are visible and close to the camera. Our subset consists of the features of 200 videos. Videos vary between 2 and 25 minutes, with a frame rate of 24 FPS, the total numbers of IPUs is 919, and of words is 62307. We shuffled all the IPUs, then split them into: training set (80\%), validation set (10\%) and test set (10\%). 
There are two test conditions: SD (Speaker Dependent) and SI (Speaker Independent). The SD condition aims to assess to what extent the model can generalize on new sentences pronounced by a speaker seen during training - training set included multiple speakers. The SI condition aims to assess the extent to which gestures predictions can be extrapolated to unseen speakers. 
Each training batch contained 128 pairs of word embeddings, F0 sequences, and their corresponding AUs, RX, RY and RZ. The loss function used is the categorical cross-entropy between predicted and actual values. We used Adam optimizer with $\beta_{1} = 0.9, \beta_{2} = 0.98$ and $\epsilon = 10^9$. We used a learning rate scheduler as in \cite{vaswani2017attention}, with \emph{warmup steps = 4000}. We applied a \emph{dropout = 0.1} to the output of each sub-layer of the transformer, and to the sums of the positional encodings in the Transformer encoder and decoder stacks. All features values were normalized between 0 and 1. The total number of the model's parameters is  2051133. 
\subsection{Objective Evaluation}
To assess the quality of the generated gestures, we used the following measures: \textbf{\emph{Root Mean Squared Error} (\emph{RMSE})}, \textbf{\emph{Pearson} Correlation Coefficient (\emph{PCC})}, \textbf{\emph{Activity Hit Ratio} (\emph{AHR})} and \textbf{\emph{Non-Activity Hit Ratio} (\emph{NAHR})}. AHR and NAHR were proposed by Freeman et al. \cite{ong2017real}, to evaluate the performance of Voice Activity Detector (VAD) systems. We also considered them since evaluating AU activity looks similar to VAD evaluation. We considered an AU as \textbf{\emph{``Activated''}} when its value is greater than \textbf{\emph{0.5}}, otherwise it is \textbf{\emph{``Not-Activated''}}. \textbf{\emph{AHR}} is the percentage of predicted AU activation with respect to ground truth. If it is greater than 100\%, it means that the model is predicting more activation than the amount of activation that is in the ground truth. \textbf{\emph{NAHR}} is the same but for non-activity. We assessed the full model using the \emph{\textbf{SD}} test set. We additionally evaluated its capacity to generalize on the \emph{\textbf{SI}} set. To evaluate the different parts of our architecture, we conducted an ablation study as follows: \emph{\textbf{(1)}} speech ablation, \emph{\textbf{(2)}} text ablation, \emph{\textbf{(3)}} CMAM ablation, and \emph{\textbf{(4)}} AU/R decoder ablation. The ablation study was evaluated using RMSE metric. 

\begin{table*}[ht]
\centering
\begin{tabular}{ccccc|cccc}
\cline{2-9}
& \multicolumn{4}{c|}{\textbf{Proposed Transformer Mode (SD)}}                                                        & \multicolumn{4}{c}{\textbf{SOTA Baseline (SD)}}                                                            \\ \cline{2-9} 
\textbf{}                           & \multicolumn{1}{c|}{\textbf{RMSE}} & \multicolumn{1}{c|}{\textbf{PCC}} & \multicolumn{1}{c|}{\textbf{AHR}} & \textbf{NAHR} & \multicolumn{1}{c|}{\textbf{RMSE}} & \multicolumn{1}{c|}{\textbf{PCC}} & \multicolumn{1}{c|}{\textbf{AHR}} & \textbf{NAHR} \\ \hline \hline
\multicolumn{1}{c|}{\textbf{AU01}} & \multicolumn{1}{c|}{0.082}        & \multicolumn{1}{c|}{1.0}          & \multicolumn{1}{c|}{98.0}         & 101.0         & \multicolumn{1}{c|}{0.20}        & \multicolumn{1}{c|}{-0.012}      & \multicolumn{1}{c|}{42.56}       & 115.95        \\ 
\multicolumn{1}{c|}{\textbf{AU02}} & \multicolumn{1}{c|}{0.028}        & \multicolumn{1}{c|}{0.99}        & \multicolumn{1}{c|}{100.0}        & 100.0         & \multicolumn{1}{c|}{0.48}        & \multicolumn{1}{c|}{-0.002}      & \multicolumn{1}{c|}{34.15}       & 107.08       \\ 
\multicolumn{1}{c|}{\textbf{AU04}} & \multicolumn{1}{c|}{0.037}        & \multicolumn{1}{c|}{1.0}          & \multicolumn{1}{c|}{99.0}         & 130.0         & \multicolumn{1}{c|}{0.53}        & \multicolumn{1}{c|}{-0.012}      & \multicolumn{1}{c|}{60.31}       & 121.85       \\ 
\multicolumn{1}{c|}{\textbf{AU05}} & \multicolumn{1}{c|}{0.023}        & \multicolumn{1}{c|}{1.00}        & \multicolumn{1}{c|}{100.0}        & 100.0         & \multicolumn{1}{c|}{0.50}        & \multicolumn{1}{c|}{-0.011}      & \multicolumn{1}{c|}{21.12}       & 135.95        \\ 
\multicolumn{1}{c|}{\textbf{AU06}} & \multicolumn{1}{c|}{0.060}        & \multicolumn{1}{c|}{1.0}          & \multicolumn{1}{c|}{99.3}         & 102.6         & \multicolumn{1}{c|}{0.48}        & \multicolumn{1}{c|}{-0.002}      & \multicolumn{1}{c|}{33.11}       & 135.05        \\ 
\multicolumn{1}{c|}{\textbf{AU07}} & \multicolumn{1}{c|}{0.10}        & \multicolumn{1}{c|}{1.0}          & \multicolumn{1}{c|}{98.0}         & 104.1         & \multicolumn{1}{c|}{0.33}        & \multicolumn{1}{c|}{-0.053}      & \multicolumn{1}{c|}{20.21}       & 131.52        \\ 
\multicolumn{1}{c|}{\textbf{RX}}   & \multicolumn{1}{c|}{0.16}        & \multicolumn{1}{c|}{1.0}          & \multicolumn{1}{c|}{\textit{NA}}  & \textit{NA}   & \multicolumn{1}{c|}{0.53}        & \multicolumn{1}{c|}{0.018}       & \multicolumn{1}{c|}{\textit{NA}}  & \textit{NA}   \\ 
\multicolumn{1}{c|}{\textbf{RY}}   & \multicolumn{1}{c|}{0.24}        & \multicolumn{1}{c|}{0.99}         & \multicolumn{1}{c|}{\textit{NA}}  & \textit{NA}   & \multicolumn{1}{c|}{0.97}        & \multicolumn{1}{c|}{-0.024}      & \multicolumn{1}{c|}{\textit{NA}}  & \textit{NA}   \\ 
\multicolumn{1}{c|}{\textbf{RZ}}   & \multicolumn{1}{c|}{0.30}        & \multicolumn{1}{c|}{0.94}         & \multicolumn{1}{c|}{\textit{NA}}  & \textit{NA}   & \multicolumn{1}{c|}{0.22}        & \multicolumn{1}{c|}{0.003}       & \multicolumn{1}{c|}{\textit{NA}}  & \textit{NA}   \\ \hline
\end{tabular}%
\caption{Objective Evaluation: comparison of proposed transformer model vs. SOTA baseline model}
\label{tab:my-table}
\vspace{-0.5cm}
\end{table*}

\subsection{Comparing to LSTM-based Baseline Model}
We compared our approach to a sequence to sequence LSTM-based model \cite{fares2020towards} to predict upper-face movements based on speech and text. This LSTM model\cite{fares2020towards} was developed only for predicting eyebrows movements \emph{\textbf{(AUs)}}. We extended it to include head movements \emph{\textbf{(R)}}. This extended model consists of mapping sequences of F0 and the Bert embedding that correspond to a Word \emph{\textbf{(W)}}, to the sequences of \emph{\textbf{(AUs/R)}} of that corresponding \emph{\textbf{W}}. It employs 2 layers of Bidirectional LSTMs to encode the concatenation of word-level F0 contours and Bert embeddings. The hidden internal states are then transmitted to 6 decoders to produce the corresponding AUs and R for a given input. The decoders are followed by \emph{Dense Layers} with \emph{Softmax Activation}. The hyperparameters of the extended LSTM model are as follows: in both encoder and decoders, the first layer of bidirectional LSTM has 200 units, and the second one has 100 units. The activation function used in these layers is \emph{LeakyReLU} with \emph{alpha = 0.01}. We trained and tested this model using the same training, validation and test sets described in Section III - A. We trained it on 300 epochs, using a \emph{batch size = 128}. We used \emph{Root Mean Squared Propagation (RMSProp)} optimizer, and \emph{Categorical Cross Entropy Loss}. 
We used the same metrics described in III.B to compare this SOTA Baseline model to our Transformer-based model.

\subsection{Subjective Evaluation}
To investigate human perception of the facial gestures produced by our model, we conducted two different experimental studies using the virtual agent Greta \cite{pelachaud2017greta}. We followed the recommendations proposed in \cite{wolfert2021review}, by adapting them to facial gesture generation and assessed the \emph{naturalness}, \emph{coherence}, and \emph{human-likeness} of the virtual agent's gestures. Since we are not evaluating deictic and iconic gestures, we did not use the metrics \emph{appropriateness}, and \emph{intelligibility} as proposed in \cite{wolfert2021review}. We added the metrics \emph{synchronization}, and \emph{alignment} to evaluate the gestures' temporal property with speech. Participants in both studies were fluent in English, with a University degree, and recruited on Prolific, a crowd sourcing website. We added attention checks at the beginning of our perceptual evaluations, to filter out inattentive participants. The first study was done by 35 participants, and consisted of presenting 16 videos: each video showed the virtual agent saying a sequence of words that corresponds to a sequence of IPUs. We considered 4 conditions: 4 videos (condition \textbf{M}) used our full model of \emph{\textbf{SD}} gestures predictions; 4  videos (condition \textbf{GT}) were simulated using the gestures extracted from TED videos, which serve as ground truth; 4 videos of the virtual agent were simulated using the LSTM-based Baseline model of \emph{\textbf{SD}} predictions (condition \textbf{SOTA}). The remaining 4 videos were produced using predicted gesture animation of IPUs with the sound of other IPUs (condition \textbf{E}). 
The second study was conducted by 55 participants. The goal of the second study was to evaluate our model when simulated with \textbf{SI} data, and therefore its capability to generalise to new speakers. It included 8 videos: 4 were simulated with our model's \textbf{SI} predictions, and 4 using \textbf{SI} gestures extracted from \emph{\textbf{SI}} set (described in \emph{III.A}), which serve as ground truth. For each video in both studies, participants were asked to rate the 5 factors, namely \emph{naturalness}, \emph{coherence}, \emph{human-likeness}, \emph{synchronization}, and \emph{alignment} of the virtual agent's gestures on a 1 to 7 likert scale \cite{wolfert2021review}. The questions were listed in a random order. The agent's 
mouth movements 
were blurred to prevent participants from getting distracted by these gestures which were not inferred by our model, and therefore focus on the model's generated gestures.

\section{Results and Discussion}
Table \ref{tab:my-table} reports the model's as well as the \textbf{SOTA}'s objective evaluation results using the same \emph{\textbf{SD}} test set. Results reveal that RMSE errors are much smaller for \textbf{M} ($0.0229\le$\emph{error}$\leq0.3$) than \textbf{SOTA} ($0.2046\le$\emph{error}$\leq$0.9786). On the other hand, PCC coefficients show that \textbf{M}'s predictions (0.94$\leq$\emph{PCC}$\leq$1.0) are more correlated than \textbf{SOTA}'s predictions (-0.0525$\leq$\emph{PCC}$\leq$0.0179) to \textbf{GT}. AHR and NAHR were calculated to measure the activation of AUs only, since they are not applicable for R. Results show that \textbf{M} predicts better the activation rate AHR (\emph{AHR}$\ge$98.0) than \textbf{SOTA} (20.211$\leq$\emph{AHR}$\leq$60.314). The non-activation rate is higher for \textbf{SOTA} (107.084$\leq$\emph{NAHR}$\leq$135.95) than for \textbf{M}(100$\leq$\emph{NAHR}$\leq$130). This constitutes objective validation that \textbf{M} gives better results than \textbf{SOTA} in terms of \emph{error}, \emph{correlation}, and AU's \emph{activation rate}.
 Ablation studies conducted on our model resulted in higher RMSE errors when performing Speech and Text Ablation for some AUs/R. For instance, AU01 RMSE error (0.0819 for full model) increased to 0.0836 with text ablation and to 0.09 with speech ablation. On the contrary, RZ RMSE score (0.3036 for full model) increased to 0.3089 with text ablation, and 0.3111 with speech ablation. 
AU/R Decoder ablation resulted in even higher RMSE errors especially for head rotations. AUs and R RMSE scores increased after CMAM ablation (i.e. AU01 RMSE increased to 0.0901). This constitutes objective validation that the use of multi-modal inputs (speech and text modalities) in \textbf{M} improves predictions. Thus we can also conclude that CMAM module is an efficient and a key component of our model, as it improves the generation accuracy of face gestures and head rotations.
As mentioned previously, we also tested our model on the \emph{\textbf{SI}} set. RMSE errors are between 0.301 and 0.89 for AUs, and between 0.25 and 0.93 for R. As we could expect, we got higher errors than the errors we had for the \emph{\textbf{SD}} condition (Table \ref{tab:my-table}) since the speakers in \emph{\textbf{SI}} set were not seen by our model during the training phase.  

\begin{figure*}[!h]
    \centering
    \subfloat[][\centering Speaker Dependent (SD)]{{\includegraphics[width=8.5cm]{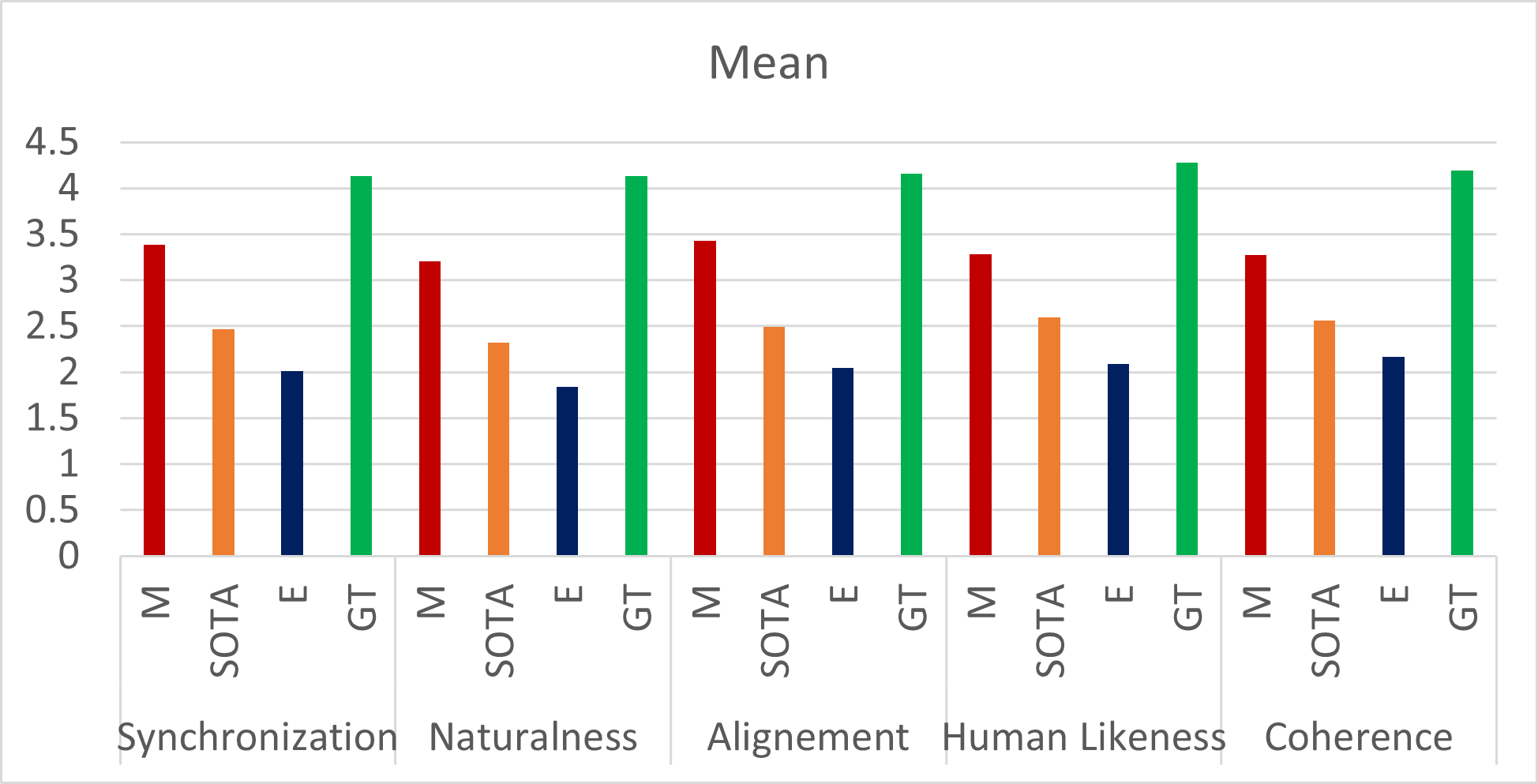} }}%
    \label{fig:SubjectiveEval_SD}
    \qquad
    \subfloat[][\centering Speaker Independent (SI)]{{\includegraphics[width=8.5cm]{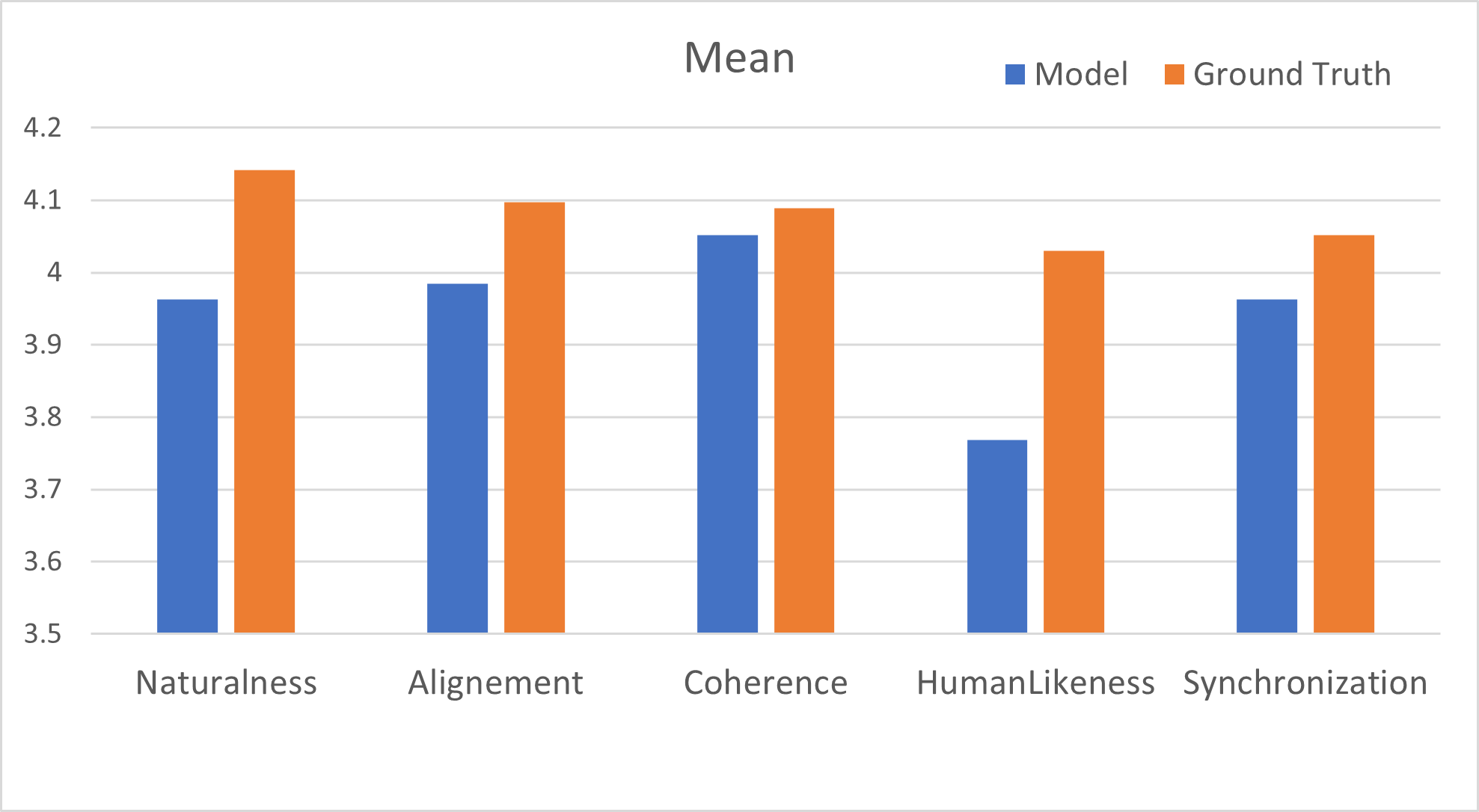} }}%
    \label{fig:SubjectiveEval_SI}
    \caption{Subjective Evaluation Results}%
    \label{fig:SubjEval}%
    \vspace{-0.2cm}
\end{figure*}

For our first perceptive study conducted on \emph{\textbf{SD}} data, Fig. \ref{fig:SubjEval} (a) shows the mean scores obtained on the 5 factors for the 4 conditions: our  model \textbf{M}, the baseline \textbf{SOTA},  the ground truth \textbf{GT}, and the error \textbf{E}. For the 5 factors, \textbf{M} is perceived as much closer to \textbf{GT} than \textbf{SOTA} and \textbf{E}, especially in terms of  \emph{Alignment} and \emph{Synchronization} between speech and gestures. The mean difference between \textbf{M} and \textbf{GT} is 0.72 for \emph{Alignment} and 0.74 for \emph{Synchronization} (Fig.\ref{fig:SubjEval} (a)). \textbf{SOTA} is perceived as the closest to \textbf{E} especially in terms of \emph{Coherence}, and \emph{Alignment}: the mean difference is 0.4 and 0.44 respectively (Fig.\ref{fig:SubjEval} (a)).  
We also conducted \emph{ANOVA Between-Subjects} to test whether there are significant differences between the 4 conditions along all the 5 factors. Results showed that for all factors, tests were significant (\emph{\textbf{p}}$<$0.001). We additionally performed a post-hoc \emph{Fisher’s LSD} Test to do pair-wise comparisons of the means between the factors of all conditions. 
Significant results (\emph{\textbf{p}}$<$0.001) were found for all the factors when comparing the condition \textbf{GT} with the 3 other conditions and when comparing our model \textbf{M} with \textbf{SOTA} and \textbf{E}. In particular, there were significant differences between \textbf{M} and \textbf{SOTA} in terms of the 5 factors (\emph{\textbf{p}}$<$0.007). 
This constitutes experimental validation that when used with \emph{\textbf{SD}} data, condition \textbf{M} is perceived significantly better than \textbf{SOTA} and \textbf{E} for all the factors. The difference for the 5 factors between \textbf{M} and \textbf{GT} is not significant. 
For our second perceptive study conducted on \emph{\textbf{SI}} data, Fig. \ref{fig:SubjEval} (b) shows the means scores obtained on all factors for conditions \textbf{M} and \textbf{GT}: for the 5 factors \textbf{M} is perceived as close to \textbf{GT}. 
As our data were not normally distributed (Shapiro test’s p$<$0.5), we conducted a post-hoc unpaired Wilcoxon test on
each factor for the two conditions. For all tests, we could not find significant differences (p$>$0.05). Thus our model when used with \emph{\textbf{SI}} data and \textbf{GT} received similar values for the 5 factors. 

\section{Conclusion}
We have presented a new approach for modelling upper facial and head gestures using Transformer model. We conducted objective and subjective evaluations that showed that our model produces animations that are close to the ground truth in term of expressivity while ensuring that speech and computed gestures are aligned and synchronized. In a next future, we plan to expand our model to learn different speaker styles for gesture generation.
\vspace{-0.05cm}
\bibliographystyle{IEEEbib}
\bibliography{references}
\end{document}